\documentclass[11pt]{article}

\usepackage[]{graphicx}
\usepackage{amsfonts,amsmath,amssymb}
\usepackage[numbers]{natbib}
\usepackage{times}

\newcommand{\Reyn}{\mbox{Re}}
\newcommand{\St}{\mbox{St}}
\newcommand{\Li}{\mbox{Li}}
\newcommand{\SR}{S/h^2}
\newcommand{\Juv}{$^j$}

\newcommand{\mi}{$^<$}
\newcommand{\ma}{$^>$}

\newcommand{\q}{$\hspace{8.7pt}$}
\newcommand{\xd}{\dot{x}}
\newcommand{\yd}{\dot{y}}
\newcommand{\xp}{{x'}}
\newcommand{\yp}{{y'}}
\newcommand{\A}{\theta_0}

\newcommand{\etal}{\textit{et al. }}

\title{\bf Optimal Strouhal number for swimming animals}

\author{Christophe Eloy\footnote{Christophe.Eloy@irphe.univ-mrs.fr}
\\
\normalsize{\textit{Department of Mechanical and Aerospace Engineering,}}\\
\normalsize{\textit{University of California, San Diego, La Jolla, CA 92093, USA}}\\
\normalsize{\textit{IRPHE, CNRS \& Aix-Marseille Universit\'e, 49 rue Joliot-Curie, 13013 Marseille, France}}
}
\date{}

\begin{document}
\baselineskip12pt

\maketitle

\begin{quote}
To evaluate the swimming performances of aquatic animals, an important dimensionless quantity is the Strouhal number, $\St = fA/U$, with $f$ the tail-beat frequency, $A$ the peak-to-peak tail amplitude, and $U$ the swimming velocity. Experiments with flapping foils have exhibited maximum propulsive efficiency in the interval $0.25 < \St < 0.35$ and it has been argued that animals likely evolved to swim in the same narrow interval. Using Lighthill's elongated-body theory to address undulatory propulsion, it is demonstrated here that the optimal Strouhal number increases from $0.15$ to $0.8$ for animals spanning from the largest cetaceans to the smallest tadpoles. To assess the validity of this model, the swimming kinematics of $53$ different species of aquatic animals have been compiled from the literature and it shows that their Strouhal numbers are consistently near the predicted optimum.
\end{quote}

\begin{quote}
\end{quote}

\section{Introduction}
\subsection{Strouhal number}

In 1915, Lord Rayleigh published a paper on the `principle of similitude' \cite{Rayleigh1915} which served as an incentive for modern dimensional analysis \cite{Rott1992}.  
Out of the many examples given in this paper is the {\AE}olian harp, an instrument that produces musical sound when wind blows across its strings. From similarity, Rayleigh showed that the sound frequency $f$ should be equal to the ratio of wind velocity $U$ to the diameter of the string $d$ multiplied by a function of the Reynolds number, in agreement with experimental observations of Strouhal \cite{Strouhal1878}. 
Later the same year \cite{Rayleigh1915a}, Lord Rayleigh introduced what he called the Strouhal number, whose definition eventually changed to be the inverse of Rayleigh's suggestion: $\St=fd/U$ \cite{Benard1926}. Since then, the Strouhal number has been used extensively to measure in an appropriate dimensionless manner the frequency  of vortex shedding behind bluff bodies. For a circular cylinder or a sphere, it is found to be approximately equal to $0.2$ over a broad range of Reynolds numbers \cite{Roshko1954,Williamson1996}.

\begin{figure}
   \centering
   \includegraphics[scale=0.5]{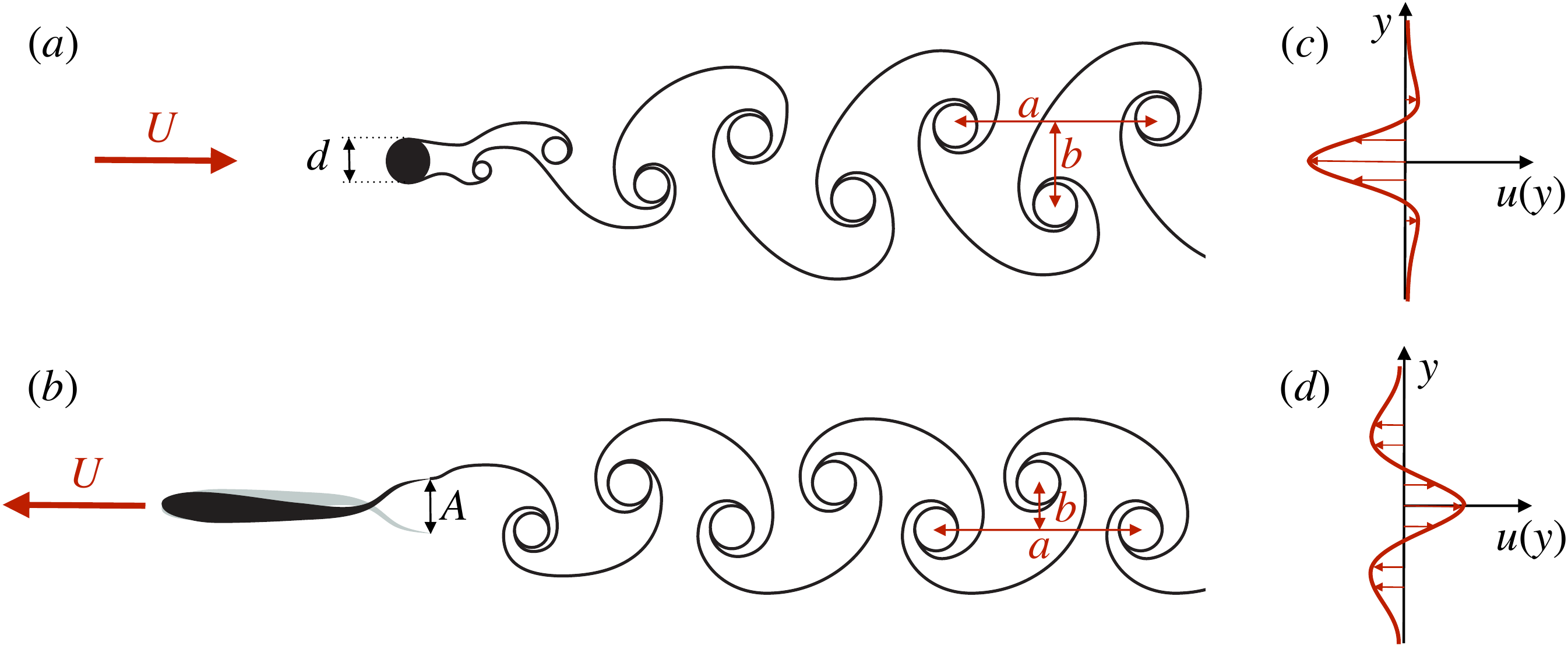} 
   \caption{Schematic view of the (\textit{a}) B\'enard-von K\'arm\'an (BvK) vortex street behind a circular cylinder and (\textit{b}) reverse B\'enard-von K\'arm\'an (rBvK) vortex street behind a swimming fish. The lines in the wakes illustrate what can be obtained typically with dye visualisations. The average perturbation flow $u(y)$ in the far wake is a jet toward the cylinder (\textit{c}) and away from the fish (\textit{d}) respectively. Both of these jets are surrounded by a region of counterflow.}
   \label{fig:wakes}
\end{figure}

The Strouhal number is intimately linked to the arrangements of vortices in the wake as already pointed out by Rayleigh \cite{Rayleigh1915a}. Von K\'arm\'an \cite{Karman1911} showed that two infinite rows of point vortices are always unstable unless their spacing ratio has a particular value $b/a=0.281$ (see figure~\ref {fig:wakes}\textit{a}). Assuming that the vortices in the wake travel at the velocity $U_w < U$, the vortex shedding frequency is then $f = U_w/a$ and the Strouhal number is linked to the spacing ratio through $\St= (b/a) (d/b) (U/U_w)$. The Strouhal number can therefore be predicted based on estimation of the spreading factor $b/d$ and the velocity ratio $U_w/U$ \cite{Roshko1954a}. 

Another approach to predict the Strouhal number consists in analysing the local stability properties of the wake \cite{Huerre1990}. To do so, a base flow is considered which can either be a steady solution of the Navier--Stokes equations around the bluff body or the time-average flow obtained through experiments or numerical simulations. 
In the near wake of this base flow, a transition from convective to absolute instability occurs \cite{Pier2002}. This region acts as a source generating disturbances advected and amplified downstream and tunes the entire wake to its frequency. The Strouhal number can thus be predicted by examining the regions of absolute and convective instability in the wake.
  
In the context of swimming, the Strouhal number has been introduced in the nineties by Triantafyllou \etal with two innovative papers \cite{Triantafyllou1993, Triantafyllou1991} (see also some recent reviews on swimming \cite{Lauder2005,Sfakiotakis1999,Triantafyllou2000}). It is defined as
\begin{equation}
\label{eq:St}
\St=\frac{fA}{U},
\end{equation}
where $f$ is the tail-beat frequency, $A$ is the peak-to-peak amplitude at the tail tip and $U$ is the average swimming speed. The argument of Triantafyllou \etal \cite{Triantafyllou1993, Triantafyllou1991} relies on the observation that the wake behind a swimming animal resembles the B\'enard-von K\'arm\'an (BvK) vortex street observed behind bluff bodies except that the sign of vortices are inverted giving a reverse B\'enard-von K\'arm\'an (rBvK) street (see figure~\ref {fig:wakes}\textit{b}). 

In the BvK street, the average flow exhibits a deficit of velocity compared to the imposed flow $U$, indicating that longitudinal momentum has been lost and that a drag force is exerted on the bluff body (figure~\ref {fig:wakes}\textit{c}). However, swimming animals are self-propelled and therefore no net drag nor thrust is exerted on average when they swim at constant speed: the resulting rBvK wake is therefore momentumless and exhibits on average a jet around the centerline surrounded by a region of counterflow (figure~\ref {fig:wakes}\textit{d}). Note that when the amplitude of motion $A$  is increased and the swimming velocity is held constant, there is a transition from the BvK to the rBvK street. Right at the transition, the vortices are all aligned with the swimming direction but this case still corresponds to a net drag on the body \cite{Godoy-Diana2008}.

Applying similar techniques to the ones used to study the stability of bluff body wakes, Triantafyllou \etal \cite{Triantafyllou1993, Triantafyllou1991} have shown that wakes associated to net thrust are only convectively unstable (there is no region of absolute instability). Such wakes thus acts as amplifier in a narrow range of frequencies which was found to correspond to the interval $0.25<\St<0.35$ for a family of two-dimensional wakes obtained by fitting the experimental results of Koochesfahani \cite{Koochesfahani1989}. They argued that swimming animals likely evolved to exploit this amplification to reduce the swimming costs and hence should be observed to swim in the same narrow interval of Strouhal numbers. In parallel, experiments have been carried out by the same group  \cite{Anderson1998,Read2003,Schouveiler2005,Triantafyllou1993} with rigid airfoils submitted to harmonic flapping,  confirming that maximum efficiency could be reached in the same interval.  

Note that this narrow interval might not be unrelated to the stability result of von K\'arm\'an. Because there is a mirror symmetry between an infinite BvK street and a rBvK street, rBvK wakes are also unstable unless the spacing ratio is $b/a=0.281$ (figure~\ref {fig:wakes}\textit{b}). Making now the reasonable assumptions that the width $b$ of the wake is equal to the amplitude of swimming  $A$ and that the vortices in the wake have no velocity such that $f=U/a$, this stable spacing ratio corresponds to $\St=b/a=0.281$. This value is in the interval proposed by  Triantafyllou \etal \cite{Triantafyllou1993, Triantafyllou1991} but the precise link between these two approaches has yet to be understood. 

In their papers,  Triantafyllou \etal \cite{Triantafyllou1993, Triantafyllou1991} analysed twelve species (dolphins, sharks, some scombroids and other bony fishes) whose swimming kinematics were found in the literature  and concluded that most of these swimming animals indeed swim in the interval $0.25<\St<0.35$. More recently, Taylor \etal \cite{Taylor2003} have shown that birds, bats and insects in cruising flight flap their wings within a similar narrow range of Strouhal number, $0.2<\St<0.4$. To explain this apparent universal range of Strouhal number observed in nature, they argued that animals tune their frequency to achieve maximum propulsive efficiency.  

In this work, the optimal Strouhal number for swimming will be addressed with a different approach. First, Lighthill's large-amplitude elongated-body theory will be introduced and discussed. Within this theoretical framework, it will be shown in \S 2 that an optimal Strouhal number can be calculated through a constrained minimisation problem. In \S 3, this optimal Strouhal number will be compared to the observations on different species of swimming animals found in the literature. Finally, these results will be discussed and related to the characteristics of the  wake in \S 4.  

\subsection{Lighthill's elongated-body theory}


\begin{figure}
   \centering
   \includegraphics[scale=0.5]{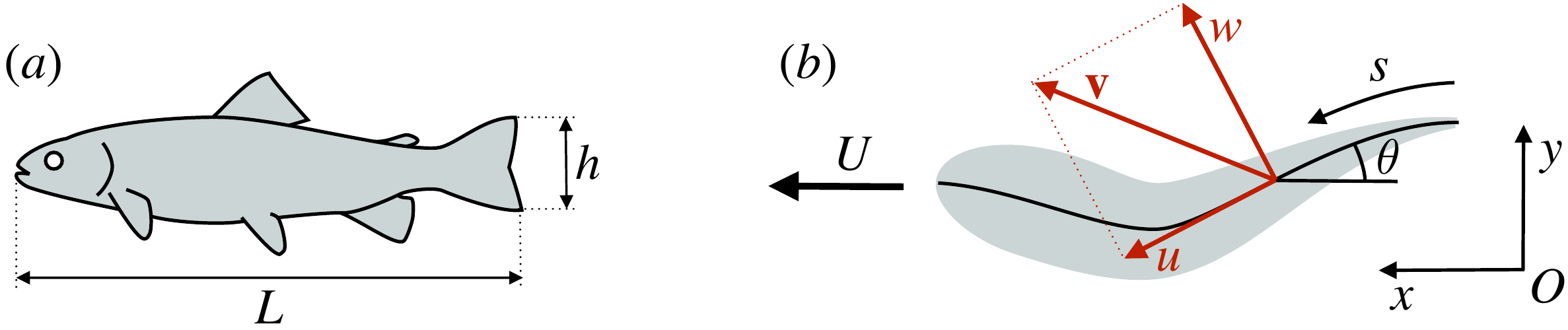} 
   \caption{(\textit{a}) Dimensions considered for the swimming animals and (\textit{b}) sketch of the problem.}
   \label{fig:sketch}
\end{figure}
Consider an aquatic animal of length $L$ swimming with constant mean velocity $U$ in the $x$-direction (see figure~\ref{fig:sketch}). The position of its body at time $t$ is described by the position $(x,y)$ of any point of the backbone. The plane $Oxy$ would be the horizontal plane for fishes and the vertical plane for cetaceans. 
Defining the curvilinear coordinate $s$ as the distance from the tail tip when the animal is straight, the functions $x(s,t)$ and $y(s,t)$ therefore describe the kinematics of swimming. 
The velocity of any point on the backbone is the time-derivative of the position, $\mathbf{v}=(\xd,\yd)$, which can be decomposed into tangential and normal velocities (see figure~\ref{fig:sketch})
\begin{subequations}
\label{eq:uw}
\begin{eqnarray}
	u & = & \xd\xp + \yd\yp , \\
	w & = & \yd\xp - \xd\yp ,
\end{eqnarray}
\end{subequations}
where the prime and dot notations refer to the derivates with respect to $s$ and $t$ respectively.

Elongated-body theory makes use of the small aspect ratio of the swimming animal. Indeed, if the aspect ratio $h/L$ is asymptotically small and if the cross section varies on a typical distance of order $L$, the forces acting on each cross sections can be assumed to be the same as those acting on an infinite cylinder with same cross section and moving with the same velocity $(u,v)$ with respect to the fluid.
The main idea behind Lighthill's elongated-body theory \cite{Lighthill1971} is then to treat perpendicular motions (given by the velocity $w$) reactively and the tangential motions (given by $u$) resistively. 
The elongated-body approximation is therefore valid when the animal is elongated enough such that $h\ll L$ (see figure~\ref{fig:sketch}) and when the Reynolds number, defined as 
\begin{equation}
\Reyn=U L /\nu,
\end{equation}
with $\nu$ the kinematic viscosity, is asymptotically large (more discussion on the validity of this theoretical framework will be given below).

The origin of the reactive force is the conservation of momentum. It can be understood if one realises that, as the animal swims, a certain volume of water has to be accelerated. This means that a certain force has to be applied to the water and reactively, the opposite force applies to the animal. The reactive force has been calculated by Lighthill \cite{Lighthill1971} and its remarkable feature is that its time-average depends only on the motion of the tail. Therefore the motion of the rest of the body does not have to be known. The same is true for the kinetic energy given to the fluid per unit time which is the only source of power loss in the elongated-body approximation. 

Following Lighthill \cite{Lighthill1971}, the mean thrust $\langle T \rangle$ (which is the reactive force on the animal projected on the $x$-direction) and the power lost in the wake $\langle E \rangle$ are given by
\begin{subequations}
\label{eq:TE}
\begin{eqnarray}
	\langle T \rangle & = & \langle m\left[w\left( \yd -{\textstyle\frac{1}{2}}w\xp\right) \right]_{s=0}\rangle , \\
	\langle E \rangle & = & \langle {\textstyle\frac{1}{2}}m\left[w^2u \right]_{s=0}\rangle ,
\end{eqnarray}
\end{subequations}
where the chevrons denote time-average, $m=\rho \pi h^2$ is the added mass per unit length at the tail tip ($s=0$) and $\rho$ is the density of water.  

In the case of steady swimming, the mean thrust $\langle T \rangle$ has to compensate the drag $D$ on the body. 
To measure the ratio between drag and available thrust, a new dimensionless number is introduced, which will be called the Lighthill number in the following 
\begin{equation}
	\Li=\frac{\pi D}{2 m U^2}=\frac{S}{h^2}C_d ,
\end{equation}
where $S$ is the total surface of the animal (or wetted surface) and $C_d$ is the drag coefficient such that $D=\frac{1}{2}\rho U^2 S C_d $.

\section{Optimisation}

\subsection{Constrained optimisation problem}
Consider now that the angle between the tail and the swimming direction is given in the vicinity of the tail tip by the harmonic motion
\begin{equation}
	\theta (s,t) = \A  \cos \left(\omega t\right), \quad \mbox{for $s\ll 1$}.
\end{equation}
Here the curvature $\theta'$ has been assumed to be zero at $s=0$ as it should be the case if one assumes that the tail is elastic and that the internal torque at the tail tip is zero. Taking the cosine and sine of $\theta$ yields $\xp$ and $\yp$, which appear as a sum of even and odd harmonics respectively
\begin{subequations}
\label{eq:xpyp}
\begin{eqnarray}
	\left[\xp\right]_{s=0} & = \left[\cos \theta\right]_{s=0} & = J_0(\A) - 2 J_2(\A) \cos \left(2\omega t\right) + \cdots, \\
	\left[\yp\right]_{s=0} & = \left[\sin \theta\right]_{s=0} & = 2 J_1(\A) \cos \left(\omega t\right) + \cdots ,
\end{eqnarray}
\end{subequations}
where $J_\nu(x)$ is the Bessel function of the first kind. The higher harmonics will be neglected in the following owing to the fact that they have a negligible influence on the final result.

To calculate the tangential and normal velocities given by (\ref{eq:uw}a,b), the functions $\xd$ and $\yd$ need to be known at the tail.
Keeping the same harmonics as in (\ref{eq:xpyp}a,b), the general form of these functions is
\begin{subequations}
\label{eq:xdyd}
\begin{eqnarray}
	\left[\xd\right]_{s=0} & = & U + \alpha U \cos (2\omega t+\phi), \\
	\left[\yd\right]_{s=0} & = & \pi\St\, U  \cos (\omega t+\psi),
\end{eqnarray}
\end{subequations}
where $\phi$ and $\psi$ are unknown phases, $\alpha$ is a dimensionless amplitude and $\St$ is the Strouhal number
given by (\ref{eq:St}).

Injecting (\ref{eq:uw}a,b), (\ref{eq:xpyp}a,b) and (\ref{eq:xdyd}a,b) into (\ref{eq:TE}a,b) and calculating the time-averages allows to express the mean thrust $\langle T \rangle$ and the mean power loss $\langle E \rangle$ as a function of the five dimensionless variables: $\A$, $\St$, $\alpha$, $\phi$ and $\psi$. 
The constrained optimisation problem then consists in solving 
\begin{equation}
\label{eq:optimisation_pb}
\min \,\langle E \rangle \quad \mbox{such that} \quad 
\left\{\begin{array}{l}
\langle T \rangle =D, \\
0 \leq (\St,\alpha) < \infty , \\
0 \leq \A \leq \pi/2, \\
0 \leq (\phi,\psi) < 2\pi.
\end{array}\right.
\end{equation}
In dimensionless form, this optimisation only depend on the Lighthill number. The problem (\ref {eq:optimisation_pb}) has been solved using the function \texttt{fmincon} in \textsc{Matlab} (The MathWorks, Inc., Natick, MA, USA). The result is a predicted optimal Strouhal number $\St(\Li)$ which is a monotonically increasing function of $\Li$ (see figure~\ref{fig:St} below).

For any value of the Lighthill number, the optimal set of dimensionless variables is always such that $\alpha$ and $\psi$ are zero, such that the functions $\xd$ and $\yd$ can be written in a simpler form for the optimal cases
\begin{subequations}
\label{eq:xdydoptimal}
\begin{eqnarray}
	\left[\xd\right]_{s=0} & = & U, \\
	\left[\yd\right]_{s=0} & = & V \left[\yp\right]_{s=0},
\end{eqnarray}
\end{subequations}
where $V$ appears as a wave speed at the tail tip and is given by
\begin{equation}
\label{eq:V}
V=\frac{\pi \St}{2J_1(\A)} U.
\end{equation}
The wave speed $V$ is always greater than the swimming speed $U$ and the ratio $U/V$ is customarily called the slip ratio. 
The fact that $\yd$ and $\yp$ are in phase in the optimal case (i.e. $\psi=0$) could have been anticipated since the same holds true in the linear limit \cite{Lighthill1970}. 
The relations (\ref {eq:xdydoptimal}a,b) mean that a simpler version of the optimisation can be performed with only two variables, $\A$ and $\St$ (or $\St$ and $U/V$ alternatively), leading to the same results.
Note also that since $\xd$ does not depend on time, the path followed by the tail tip in the frame of reference attached to the animal is a straight line in the $y$-direction. In other words, the figure of eight observed in some experiments \cite{Gray1933,Webber2001} which exists only if $\alpha \neq 0$ is not optimal within the present elongated-body framework.

\begin{figure}
   \centering
   \includegraphics[]{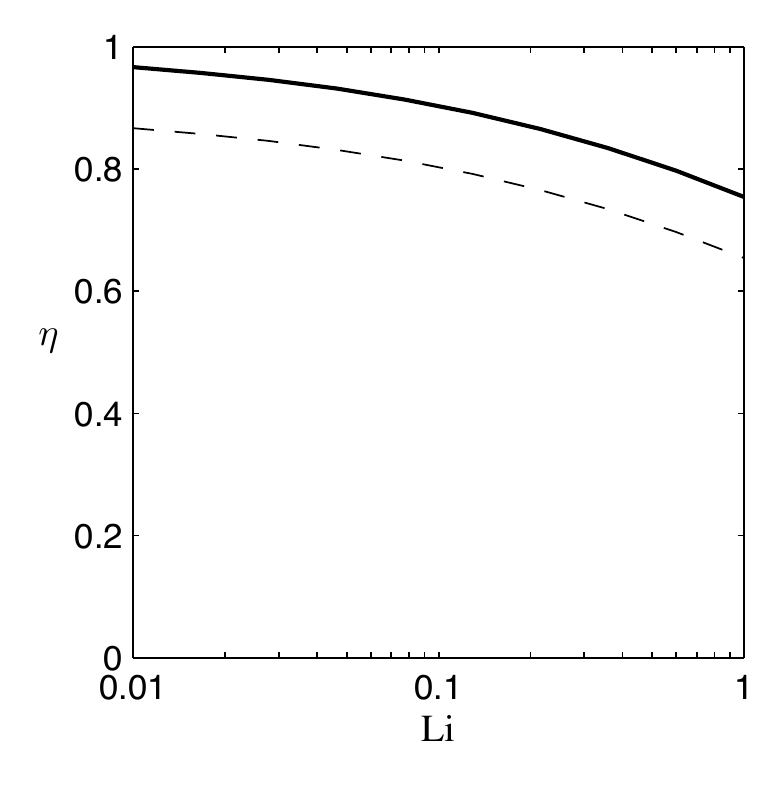} 
   \caption{Froude efficiency as a function of the Lighthill number for the optimal case (solid line) and for the acceptable range (dashed line). }
   \label{fig:eta}
\end{figure}

To estimate the range on which the Strouhal number can change without affecting appreciably the swimming performances, the Froude efficiency $\eta$ is introduced
\begin{equation}
	\eta=\frac{DU}{DU+\langle E \rangle},
\end{equation}
which expresses the ratio between the useful power $\langle T U\rangle = DU$ and the total power spent for swimming. For a given Lighthill number, the constrained optimisation yields a maximum efficiency $\eta_{\mathrm{max}} (\Li)$. In the following, any Strouhal number leading to an efficiency greater than $\eta_{\mathrm{max}}-0.1$ will be considered as acceptable (figure~\ref {fig:eta}).

\subsection{Limit of validity}
The key hypothesis of elongated-body theory is that the resistive force corresponding to perpendicular motions can be neglected. Assuming that this force acts on the whole length $L$ and that its drag coefficient is of order one (as it is the case for a cylinder), its $x$-projection scales as
\begin{equation}
F_{\mathrm{resistive}} \sim \rho h w^2 L \yp,
\end{equation}
and has to be negligible in comparison with the reactive force which scales as
\begin{equation}
F_{\mathrm{reactive}} \sim m w \yd .
\end{equation}
This is true if $(Lw)/(hV)\ll 1$ and since  $w\sim U (1 - U/V) \St$, this corresponds to
\begin{equation}
\label{eq:validity}
\frac{U}{V}\left(1-\frac{U}{V}\right) \St \ll \frac{h}{L}.
\end{equation}
This  condition will be fulfilled for most animals with a fairly wide tail. However, elongated fish such 
as eels and knife-fish, for which $h/L < 0.1$, will not in general meet this criterion. For these animals, a model taking into account the resistive normal force would be necessary. This may seem counterintuitive, but it means that Lighthill's elongated-body theory \cite{Lighthill1971} is not valid when the body is too elongated. 

The other assumption behind elongated-body theory is that the animal cross section varies on a typical length scale of order $L$. This is not true for scombrids, dolphins and sharks that share a large aspect-ratio tail. For these animals, a two-dimensional approach would be more suited to study the propulsive performance of the tail. However, the existing two-dimensional models \cite{Lighthill1970,Wu1971b} are linear and do not allow the same sort of optimisation calculation as the one presented here. 

It can be difficult to assess the validity of Lighthill's elongated-body theory because there has been no fair comparison with numerical results so far. The recent paper by Candelier \etal \cite{Candelier2011} is one notable exception.  They showed the results of numerical simulations for an eel-like swimmer with aspect ratio $h/L=0.1$, slip ratio $U/V=0.4$, Reynolds number $\Reyn=6\times 10^5$, and for two different Strouhal numbers $\St=0.2$ and $0.8$. In the first case ($\St=0.2$), the validity condition (\ref{eq:validity}) is fairly satisfied (the left-hand side being equal to $0.048$) and the thrust is well approximated by Lighthill's elongated-body theory. In the second case ($\St=0.8$), the condition (\ref{eq:validity}) is not satisfied and viscous forces play an significant role in the thrust. Note that, in these simulations, the swimmers are not self-propelled and more numerical results are clearly needed to evaluate properly Lighthill's elongated-body theory.

\section{Comparison with aquatic animals}
The present paper predicts, within the framework of elongated-body theory, the optimal Strouhal number and the maximum angle of the tail tip $\A$ as a function of the Lighthill number. To assess the validity of this model, the swimming kinematics of various species of aquatic animals have been compiled. 
Comparing these experiments and the model relies on the implicit assumption that aquatic animals swim optimally or, equivalently, that that their Froude efficiencies are maximised. This assumption is far from being obvious and one could argue that some species have evolved in ecological niches where economical steady swimming is not crucial. One could also remark that Froude efficiency is only a part of the full picture: the complete energy cost has to take into account the efficiency of muscles (which depends critically on frequency and amplitude) and the losses due to the viscoelasticity of soft tissues or damping at the intervertebral joints \cite{Cheng1998,Long1992,Long2002,McMillen2006}. 

\subsection{Remarks on the experimental studies}
The comparison with aquatic animals has to cope with several limitations of the experimental methods, some of which are listed below. 
\begin{itemize}
\item In order to make a fair comparison with the present theoretical model, the measurements have to be made when the animal is swimming steadily. This is not always evident to guarantee and it can lead to large errors. Any small positive or negative acceleration can alter the results because the mean thrust is no longer equal to the drag in that case \cite{Videler1984}.

\item The drag on swimming animals has been recognised to be difficult to measure adequately \cite{Anderson2001,Fish1999,Wu2011}. It usually depends on the swimming velocity \cite{Videler1981} and can be greatly increased when the animal is close to the water surface \cite{Videler1993}. 

\item The optimality of swimming probably depends on the swimming velocity. It seems reasonable to think that, at small speeds, the efficiency is not as crucial as at large speeds. This is due to the fact that the power spent for swimming roughly scale as $U^3$. 

\item Unfortunately, in most experimental works cited here, the Strouhal number have not been calculated directly by the authors. It results that $\St$ had to be estimated from averages. When the quantities vary over large intervals, it can result in non negligible errors since the mean of a product is usually not equal to the product of the means. The same holds true for the Lighthill number. 

\item The geometrical aspects are important to determine the Lighthill number of a given animal. In particular, the wetted surface $S$ and the tail span $d$ have to be known precisely. Regrettably, these quantities have rarely been measured by the authors of experimental kinematics studies and pictures had to be used to estimate them as explained below.  

\item Most kinematics studies have been performed in water channels (also called flumes or tunnels) where the animals swim against the current imposed by the experimentalist in a given test section. This advantages of this setup are that it allows to adjust the swimming speed precisely and it guarantees the position and the direction of the animal. However, the presence of the walls can affect the swimming mode as it has been shown by Webb \cite{Webb1993}. This effect can be particularly large when the experiments are performed in respirometers where the volume of the test section has to be small enough to allow correct measurements of the variation of oxygen concentration. Moreover, in some water channels, the turbulence rate can be important and this may change the value of the drag.

\item The present theoretical analysis not only predicts the optimal Strouhal number for a given Lighthill number but also the maximum angle $\A$ at the  tail tip. However, few experimental studies report this maximal angle and thus the comparison is not very significant as it will be discussed below. Through equation (\ref {eq:V}), this maximum angle can be related to a wave speed $V$ which can be compared to the swimming speed through the slip ratio $U/V$. Contrarily to the maximum angle, the slip ratio has been widely measured in experiments. However, the wave speed usually depends on the curvilinear coordinate and it can be shown to either accelerate toward to tail tip \cite{Gillis1997} or decelerate \cite{Videler1984}. This wave speed can also be calculated through the apparent wavelength $\lambda$ of the animal deflection as $V=\lambda f$, but this method  usually gives a result different from the direct measurement \cite{Webb1984}. These limitations come from the fact that the slip ratio is non local in nature: contrarily to the Strouhal number and the maximum angle it does not only depend on measurements made at the tail tip. 
\end{itemize}

The limitations of the experimental studies listed above make the comparison with the present analysis difficult. In particular, the Lighthill number can only be estimated in most cases (if not because of the lack of geometrical measurements, because of the drag coefficient). The other important point is that the optimality of swimming can never be guaranteed. 

\subsection{Methods}

Despite the limitations listed above, most of the data available in the literature on swimming kinematics of aquatic animals have been compiled. From these sources, the Lighthill and the Strouhal numbers have been determined together with the maximum angle at the tail tip and the slip ratio $U/V$ when possible. The following methods have been used to extract the experimental data. 

When it was possible, the value chosen for the swimming velocity was 75\% of the critical velocity $U_{\mathrm{crit}}$. The critical velocity, as introduced by Brett \cite{Brett1964}, measures the maximum sustained speed for a given time (between 2 and 30 minutes depending on the authors and on the species). The reason to choose this particular value of the velocity is that it allows a large enough swimming speed (for lower speed, the swimming mode may not be optimal) without being too close to the critical value where data is usually lacking. 

To evaluate the tail span of a given species, the following rules have been used. When available, the data found in the source papers have been used (using the value given by the authors when present, or measuring them from the photographs or drawings found in the article). Otherwise, pictures have been collected on Internet and used to estimate the ratio of tail span to body length. For species with no marked tail (such as eels, leeches, or crocodiles), the maximum value of the animal span in its posterior half have been used in place of the tail span. 
To estimate the wetted surface of each species, we mostly used rough estimates because this data was rarely given in the source papers. These estimates have been assumed to vary between $S=0.15L^2$ for the European eel and $0.5L^2$ for the Florida manatee, but for most fishes (not elongated ones), the value $0.4L^2$ has been used.  Note that the tail span need to be estimated with a greater precision than the wetted surface since it appears squared in the Lighthill number.

As pointed out by   different authors (see the recent review by Wu \cite{Wu2011}, for instance), measuring the drag coefficient on swimming fishes is a difficult task and the data found in the literature are not always consistent. As a matter of fact, it may not be a well-posed problem since it is impossible to distinguish the drag from the thrust when an animal is swimming \cite{Schultz2002}. An estimate is, however, needed if one wants to use Lighthill's elongated-body theory. A simple law that compares reasonably well with available data \cite{Anderson2001,Fish1999,Lighthill1971,Videler1981,Webb1984} is to take the double of the drag coefficient for a flat plate in laminar flow for small Reynolds number, and the turbulent drag coefficient of a flat plate for larger Reynolds number. The drag coefficient is then 
\begin{equation}
\label{eq:Cd}
	C_d (\Reyn) = \max \left( 2\times1.328\,\Reyn^{-1/2},0.072\,\Reyn^{-1/5} \right).
\end{equation}
The above law is usually an underestimate of the drag coefficient, particularly for animals with relatively poor streamlining. For instance, the lake sturgeon may have a drag coefficient 3.5 times that of trout of similar size probably because of its large scutes \cite{Webb1986}. Note that, when calculating the Reynolds number, one has to take into account that the dynamic viscosity varies substantially with temperature. 
The relation (\ref{eq:Cd}) has been used to calculate the drag coefficient for all animals except some mammals (beluga, bottlenose dolphin, false killer whale, harp seals, killer whale and ringed seals) for which the authors provided a drag coefficient corrected for surface effects \cite{Fish1998,Fish1988}. For the different morphotypes of goldfish studied by Blake \etal \cite{Blake2009}, the drag coefficient has also been calculated from the data fit given by the authors because these morphotypes have been selected artificially for aesthetic reasons, and as a result have a relatively large drag. 

As pointed out above, most authors did not calculate the Strouhal number for each swimming event, mostly because it was not the principal goal of their studies. Thus the Strouhal number had to be calculated using the average tail amplitude $\langle A \rangle$, the average frequency $\langle f \rangle$ and the average swimming velocity $\langle U \rangle$ for a given series of experiments, leading to potential errors. However, when they were available, the data of single events have been extracted and the Strouhal number have been calculated for each of them before the average was performed   
\cite{Archer1989,DAout1997,Gillis1997,Hess1984,Videler1984,Videler1993,Webb1973}. In other cases, the data of several sets of experiments (for different size groups, for instance) have been averaged  \cite{Dewar1994,Jayne1995,Lowe1996,Seebacher2003}.
 
From the literature, 89 different swimming kinematics have been identified. After analysis, 23 of these data have been discarded, either because better data were available for the same or a similar species, because the quality of the  data was doubtful (when it was based on a single experiment, for instance) or because the validity of elongated-body theory  as defined by equation (\ref {eq:validity}) was not ensured (this is mainly why there is no snake and no larva in the data set). 
 
The remaining 66 data represents 53 different species which have been divided in 7 different groups (table~1--3): 
8 different species of mammals, 
4 of sharks, 
8 of scombrids (a family which includes tunas, bonitos and mackerels), 
11 of fishes from the order of Perciformes and Salmoniformes  (excluding the family of scombrids), 
19 of fishes from other families (including Cypriniformes, Gadiformes and Mugiliformes), 
10 of `elongated' fishes (including eels, needlefish of the family of Belonidae, and other fishes with surface ratio $\SR$ greater than 17) and 
6 species categorised as 'others' gathering one reptile (crocodile), two frog tadpoles, two amphibians (axolotl and siren) and one annelid (leech). 

\subsection{Results}
\begin{figure}
   \centering
   \includegraphics[]{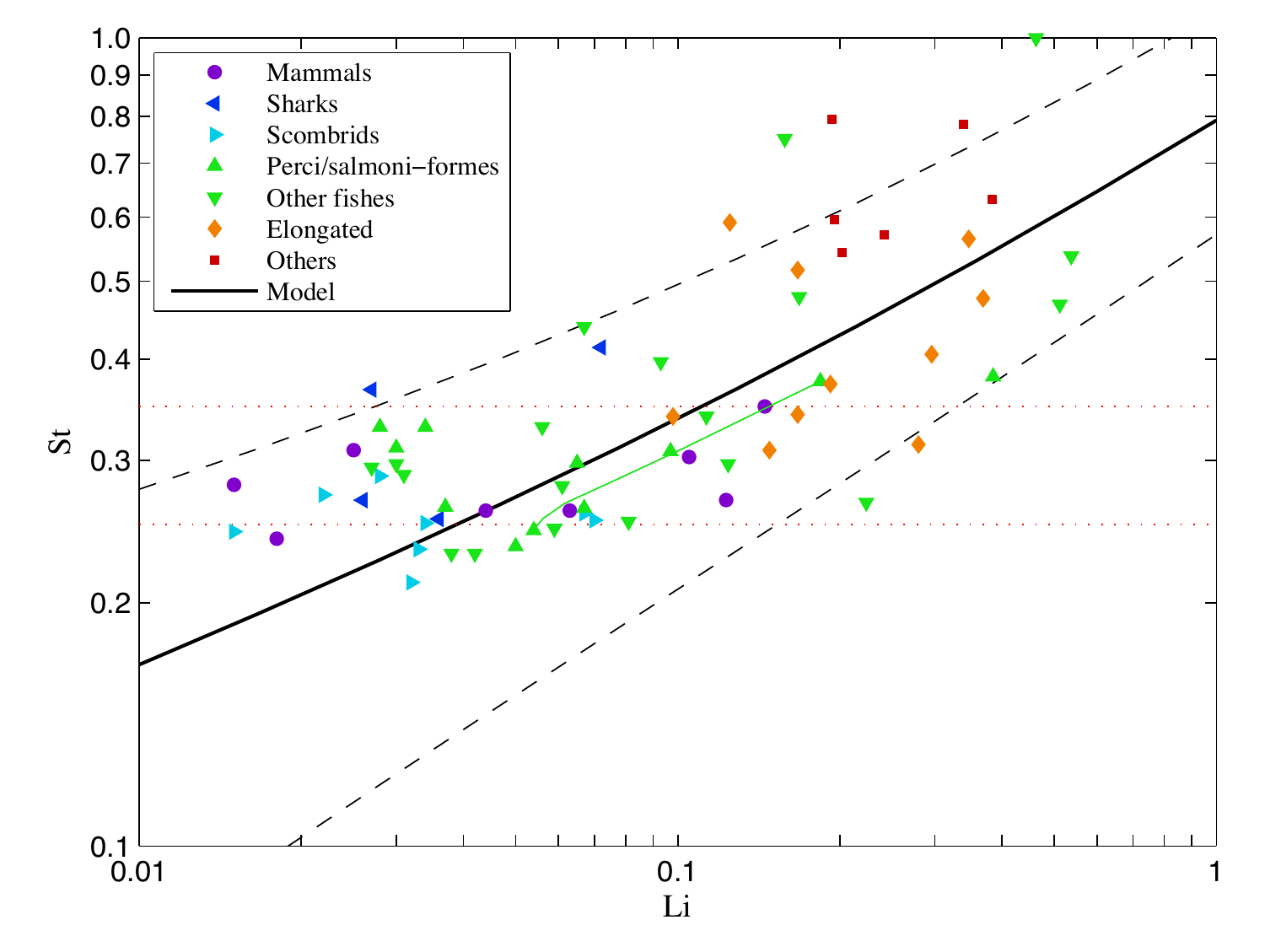} 
   \caption{Strouhal number of 53 different species of aquatic animals as a function of the Lighthill number. These animals are divided in different categories corresponding to the different symbols displayed  in the legend. The solid line is the predicted optimal Strouhal number and the dashed line correspond to the interval for which efficiency is larger than $\eta_{\mathrm{max}}-0.1$. The horizontal dotted lines correspond to the interval $0.25<\St<0.35$ suggested by Triantafyllou \etal  \cite{Triantafyllou1993}. }
   \label{fig:St}
\end{figure}

\begin{figure}
   \centering
   \includegraphics[scale=0.5]{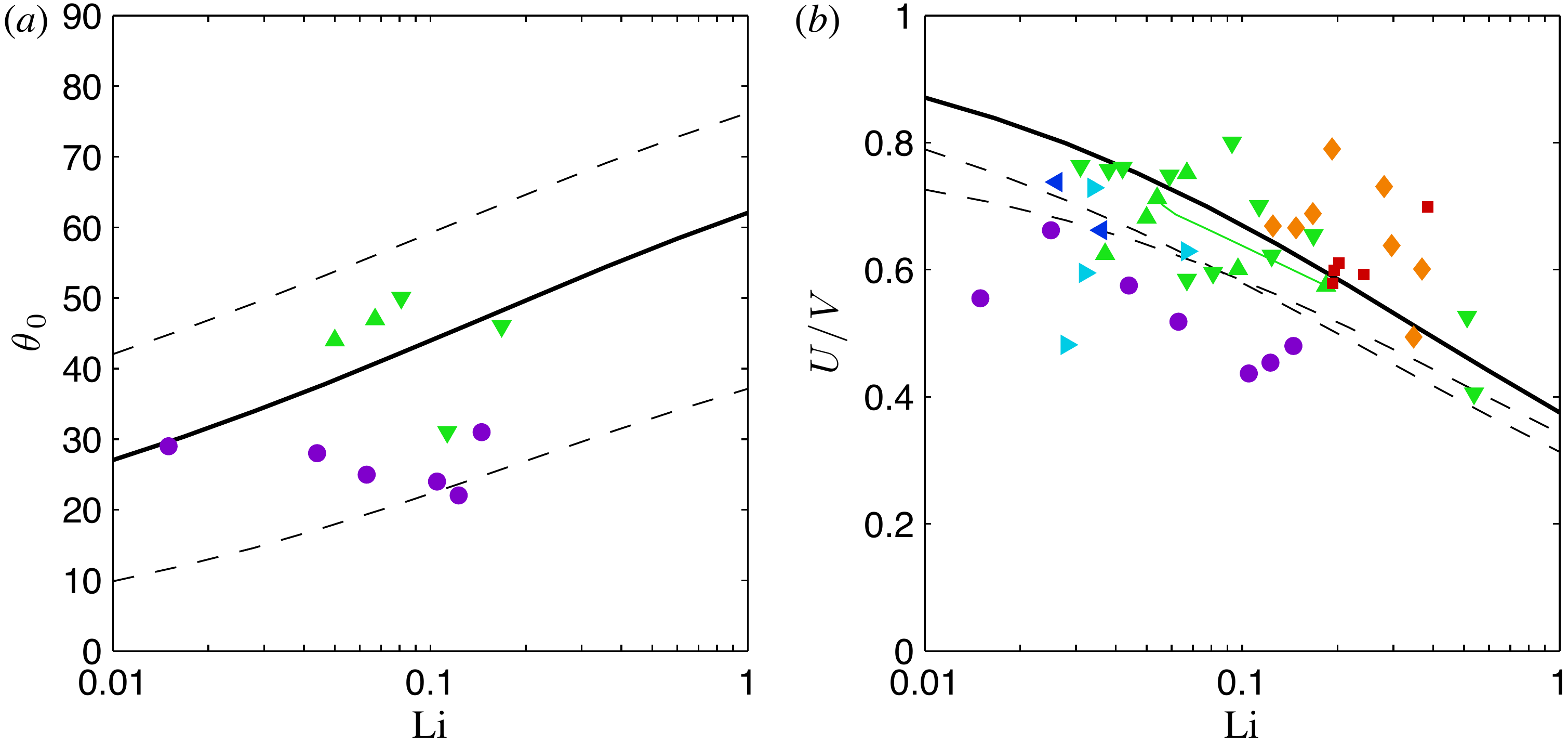} 
   \caption{(\textit{a}) Maximum angle at the tail tip $\A$ and (\textit{b}) slip ratio $U/V$  as a function of the Lighthill number (same legend as in figure~\ref {fig:St}).}
   \label{fig:A_slip}
\end{figure}

The Strouhal number, the maximum angle at the tail tip and the slip ratio predicted by the present theoretical model are compared to the observations on the different species in figures~\ref{fig:St}--\ref{fig:A_slip}. In these figures, the thick line correspond to the optimal case with Froude efficiency $\eta_{\mathrm{max}}$, and the dashed lines correspond to the interval for which the Froude efficiency is $\eta>\eta_{\mathrm{max}}-0.1$ (figure~\ref{fig:eta}). It corresponds to the acceptable range defined above, for which efficiency is close to optimal. 

As seen in figure~\ref{fig:St}, the present analysis predicts that the optimal Strouhal number increases with the Lighthill number from $0.15$ for the largest cetaceans to $0.8$ for the smallest animals considered (or, more precisely, for animals with the largest Lighthill number). This optimal Strouhal number curve can be approximated by a power law: $\St \approx 0.75 \,\Li^{1/3}$. Although the experimental observations are fairly scattered, this general trend is clearly observable for all the aquatic animals, more than 85\% of data points having a Strouhal number within the acceptable range. Note that the variation of the Strouhal number for a single species (figure 11.3B in \cite{Lauder2005}) or across different species \cite{Kayan1978} had been observed previously but there was no physical explanation.

Among these data, the results of Webb \etal \cite{Webb1984} on the rainbow trout stand out. They studied animals with total length ranging from $L=5.5\,$cm to $56\,$cm and deduced from hundreds of measurements how the different geometric and kinematic quantities varies with the length and the swimming speed of the trouts. This allows, for a single species, to see how the Strouhal number varies with the Lighthill number (the green line in figure~\ref{fig:St}). Remarkably, this line is parallel (on the log-log scale figure~\ref{fig:St}) to the the theoretical prediction. 

The comparison between the predicted maximum angle at the tail tip and the experimental observations (figure~\ref{fig:A_slip}\textit{a}) is less conclusive mostly because of the lack of data and because the acceptable range is fairly large. 

The predicted slip ratio $U/V$ has also  been compared with observations on animals (figure~\ref{fig:A_slip}\textit{b}). In each group, the slip ratio is decreasing with the Lighthill number, as predicted, but the mammals and the scombrids are clearly below the prediction, while elongated fish are clearly above. This discrepancy will be discussed below. Another feature of the slip ratio is that the optimal case correspond to a maximum: for a given Lighthill number, when efficiency is lower than the optimal,  so is $U/V$. 

Note that, again, the slip ratio deduced from the results of  Webb \etal \cite{Webb1984} on the rainbow trout (the green line in figure~\ref{fig:A_slip}\textit{b}) agrees remarkably well with the present prediction. 

\section{Discussion}

In this paper, Lighthill's elongated-body theory has been used to predict the optimal Strouhal number for swimming animals. Using the elongated-body assumptions, it appeared that the optimal Strouhal number depends on a single dimensionless quantity, which has been called the Lighthill number, and which can be regarded as the ratio of the animal drag to the available thrust. Together with the optimal Strouhal number, were predicted the maximum incident angle at the tail tip and the slip ratio, which also depend uniquely on the Lighthill number. 
These theoretical predictions have been been then compared with the swimming kinematics of 53 different species of swimming animals. It appeared that the general trends predicted by the present model are recovered in the zoological data, indicating that animals generally swim near the predicted optimum. 

The validity of the elongated-body theory is limited by two geometric quantities. First, the variations of the cross-section should occur on typical scales of the order of the animal length. This is clearly not the case for animals with high aspect-ratio tails (also called lunate tails) like cetaceans, scombrids and sharks. For sharks, additional difficulty is caused by the asymmetry of the tail and one could ask whether the Strouhal number should be based on the motion of largest lobe, the smallest lobe, or an average of the two. Second, the elongated-body theory is not adapted to anguilliform animals like eels for which the tail depth is difficult to define. For these very elongated animals, the viscous drag become relatively more important and should be included in the analysis, as pointed out above. These limitations probably explain why the slip ratio for the cetaceans and the scombrids is approximately $0.2$ smaller than predicted while the elongated fishes seem to have a slip ratio larger than predicted.

As noted above, the key feature of Lighthill's elongated-body theory is that the two quantities needed to perform the optimisation, namely the average propulsive thrust and the average power loss in the wake, only depend on local quantities evaluated at the tail tip. This property has been essential in developing the present model, but a natural question would be now to ask whether the predicted optimal tail motions  are compatible with the complete kinematics of a swimming animal. In particular, it would be important to evaluate the role of recoil \cite{Webb1992}, the effect of passive elasticity of the tail and the role of the internal mechanics in general \cite{Cheng1998,Long1992,Long2002,McMillen2006}. 

Let us now examine the relation between the Strouhal number and the characteristics of the wake. First, it may be important to remind that Lighthill's elongated-body theory includes a wake behind the swimmer (see figure~7 of Candelier \etal \cite{Candelier2011}, for instance). This wake is composed of an infinitely thin sheet of vorticity left by the passage of the trailing edge in the water. Applying Kelvin's circulation theorem, this wake is found to be composed of flatten vortex rings and contains the kinetic energy given by the animal to the fluid. But, in Lighthill's elongated-body theory, the dynamical evolution of this wake is not described. The main reason for that is that it is not relevant to this theory because it has no influence on the dynamics of the swimmer. Elongated-body theory being local in nature, the forces at a given animal cross-section only depend on the motion of this cross-section and on nothing else.   
 
\begin{figure}
   \centering
   \includegraphics[scale=0.5]{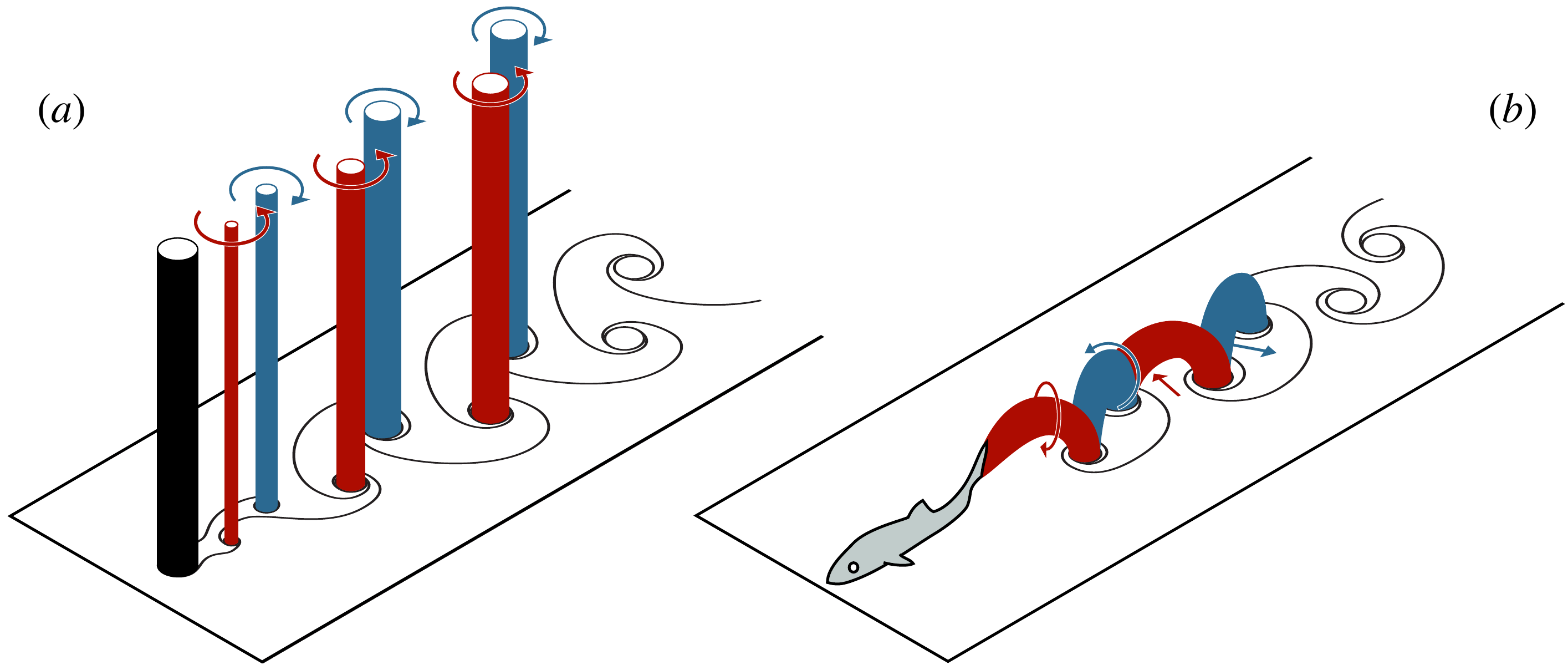} 
   \caption{Schematic three-dimensional views of the (\textit{a}) BvK and (\textit{b}) rBvK vortex streets, corresponding to the two-dimensional views of figure~\ref {fig:wakes}. The arrows indicate the rotation direction of vorticity tubes.}
   \label{fig:wakes3D}
\end{figure}
One plausible scenario for the wake is that the vorticity sheet predicted by Lighthill's elongated-body theory will eventually roll-up to form a chain of vortex rings as sketched in figure~\ref{fig:wakes3D}\textit{b}. The wake would then resemble the experimental observations of different groups \cite{Blickhan1992,Lauder2005,Muller1997,Nauen2002,Videler1993}. Another possible scenario is that, due to their self-induced velocity, these concatenated vortex rings will separate in two rows of vortex rings as observed for eels both experimentally \cite{Lauder2005,Muller2001,Tytell2004} and numerically \cite{Borazjani2008,Kern2006}. In both cases, the vortex rings are expected to  have a vertical extension approximately equal to the tail span  and a horizontal extension equal to half the stride length (invert horizontal and vertical for mammals). The stride length is here defined as $U/f$ and is  found to lie in the interval $0.5L<U/f<1.0L$ depending on the animal. Since the tail span varies in the interval $0.05L<h<0.35L$, this means that the aspect ratio of the vortex rings varies from near circular to elongated in the swimming direction (for elongated fish). Note that these differences  in aspect ratio may be one of the reason why the wakes are different for elongated fish. 

As a rule, there is a major difference between the inherently three-dimensional wake behind a swimming animal and the two-dimensional wake observed behind an infinite cylinder (as drawn in figure~\ref{fig:wakes3D}). Besides the geometry, one important difference is that the velocity field induced by a vortex line (or tube) decreases as the inverse power of the distance, while it decreases as the inverse square for a vortex ring. This means that the far wake has far less impact on the body for a three-dimensional wake, in line with Lighthill's elongated-body theory. 
The characteristics  of the wake behind a swimming animal could then just be a consequence of good propulsion and not a cause, as already suggested by  M{\"u}ller \etal \cite{Muller1997} with the idea of `fish foot prints'.

Now coming back to the results of Triantafyllou \etal \cite{Triantafyllou1993, Triantafyllou1991} on the stability and efficiency of wakes, it appears that both their theory and experiments \cite{Anderson1998,Read2003,Schouveiler2005,Triantafyllou1993} are based on two-dimensional flows. If this limit case can be suited to animals with large aspect-ratio tails (such as tunas, sharks, dolphins, etc), some more specific work seems needed to apply it to other animals. In particular, it would be interesting to understand the efficiency and stability of a three-dimensional momentumless wake. 

For animals with large aspect-ratio tails though, one striking fact is that the present model (with no modelling of the wake) and the theoretical prediction of Triantafyllou \etal \cite{Triantafyllou1993, Triantafyllou1991} (based essentially on the wake) converge to give a similar interval for the Strouhal number: roughly $0.2<\St<0.4$, for non-elongated fishes and cetaceans. One possibility is that these animals have evolved to optimise both the formation of a coherent wake and the elongated-body efficiency. This would explain why the geometrical characteristics of these animals, in particular the ratio of tail span to body length, vary so little among the species.  

~

{\small The author acknowledges support from the European Commission through a Marie Curie fellowship.}

\bibliographystyle{authordate1}
\bibliography{/Users/Ch/Documents/LaTeX/biblio}

\baselineskip11pt
\begin{table}
\label{tab:1}
\caption{
Swimming kinematics for 53 different species of mammals, fish, amphibians and reptiles (continued in tables~2-3). 
The different columns are: the animal length, $L$ in cm; the surface ratio, $\SR$; the Reynolds number, $\Reyn$;  the Lighthill number, $\Li$; the Strouhal number, $\St$, the slip ratio, 
$U/V$; and the maximum incident angle at the tail tip, $\A$ in degrees.  
The superscript $j$ marks the juvenile animals and the superscripts $<$ and $>$ indicate the minimum and maximum values of the continuous dataset obtained on the raibow trout by Webb \etal \cite{Webb1984}.}
\begin{center}
\begin{tabular}{@{}lrr cc cc cc@{}}
\hline\\[-5pt]
Species 				& $L$ (cm)	& $\SR$		& $\Reyn$			& $\Li$ 	& $\St$ 	& $U/V$		& $\A$ (deg)& Sources \\[5pt] \hline\\[-5pt]

\textbf{Mammals	}		&			&			&					&			&			&			&		& \\
Beluga					& $364$\q	& $7.3$		& $8.0\times 10^6$	& $0.145$	& $0.35$	& $0.48$	& $31$	& \cite{Fish1998,Fish1999,Rohr2004}\\
Bottlenose dolphin		& $258$\q	& $7.9$		& $1.2\times 10^7$	& $0.063$	& $0.26$	& $0.52$	& $25$	& \cite{Fish1998,Fish1999,Rohr2004}\\
False killer whale		& $379$\q	& $7.8$		& $2.1\times 10^7$	& $0.044$	& $0.26$	& $0.57$	& $28$	& \cite{Fish1998,Fish1999,Rohr2004}\\
Florida manatee			& $334$\q	& $7.5$		& $4.4\times 10^6$	& $0.025$	& $0.31$	& $0.66$	& 		& \cite{Kojeszewski2007}\\
Harp seal				& $153$\q	& $7.7$		& $1.6\times 10^6$	& $0.123$	& $0.27$	& $0.45$	& $22$	& \cite{Fish1988}\\
Killer whale			& $473$\q	& $7.3$		& $2.6\times 10^7$	& $0.015$	& $0.28$	& $0.56$	& $29$	& \cite{Fish1998,Fish1999,Rohr2004}\\
Ringed seal				& $106$\q	& $8.7$		& $1.3\times 10^6$	& $0.105$	& $0.30$	& $0.44$	& $24$	& \cite{Fish1988}\\
White-sided dolphin		& $221$\q	& $6.5$		& $1.3\times 10^7$	& $0.018$	& $0.24$	& 			& 		& \cite{Fish1998,Fish1999,Rohr2004}\\[5pt]
							
\textbf{Sharks	}		&			&			&					&			&			&			&		& \\
Blacktip reef shark		& $97$\q	& $7.6$		& $8.3\times 10^5$	& $0.036$	& $0.25$	& $0.66$	&		& \cite{Webb1982}\\
Bonnethead shark		& $93$\q	& $5.5$		& $8.0\times 10^5$	& $0.026$	& $0.27$	& $0.74$	&		& \cite{Webb1982}\\
Nurse shark				& $220$\q	& $17.8$	& $1.8\times 10^6$	& $0.072$	& $0.41$	& 			& 		& \cite{Webb1982}\\
Scalloped hammerhead\Juv&$59$\q	& $4.9$	& $3.8\times 10^5$	& $0.027$	& $0.37$	& 			&		& \cite{Lowe1996}\\[5pt]
							
\textbf{Scombrids}		&			&			&					&			&			&			&		& \\
Atlantic mackerel		& $32$\q	& $6.7$		& $5.8\times 10^5$	& $0.034$	& $0.25$	& $0.73$	&		& \cite{Videler1984}\\
Chub mackerel 			& $21$\q	& $10.5$	& $1.6\times 10^5$	& $0.070$	& $0.25$	& 			&		& \cite{Dickson2002}\\
Chub mackerel\Juv 		& $21$\q	& $10.5$	& $1.8\times 10^5$	& $0.067$	& $0.26$	& $0.63$	&		& \cite{Donley2000}\\
Giant bluefin tuna		& $250$\q	& $4.8$		& $5.7\times 10^6$	& $0.015$	& $0.24$	& 			&		& \cite{Wardle1989}\\
Kawakawa tuna\Juv 		& $21$\q	& $5.0$		& $1.8\times 10^5$	& $0.032$	& $0.21$	& $0.60$	&		& \cite{Donley2000}\\
Pacific bonito			& $47$\q	& $6.2$		& $4.5\times 10^5$	& $0.033$	& $0.23$	& 			&		& \cite{Dowis2003}\\
Skipjack tuna			& $57$\q	& $5.8$		& $2.2\times 10^6$	& $0.022$	& $0.27$	& 			&		& \cite{Yuen1966}\\
Yellowfin tuna			& $53$\q	& $5.5$		& $6.1\times 10^5$	& $0.028$	& $0.29$	& $0.48$	&		& \cite{Dewar1994}\\[5pt]
\end{tabular}
\end{center}
\end{table}

\begin{table}
\label{tab:2}
\caption{see Table 1.}
\begin{center}
\begin{tabular}{@{}lrr cc cc cc@{}}
\hline\\[-5pt]
Species 				& $L$ (cm)	& $\SR$		& $\Reyn$			& $\Li$ 	& $\St$ 	& $U/V$		& $\A$ (deg)& Sources \\[5pt] \hline\\[-5pt]
							
\textbf{Perci/salmoni-formes}	&	&			&					&			&			&			&		& \\	
Atlantic salmon			& $66$\q	& $6.7$		& $3.7\times 10^5$	& $0.037$	& $0.26$	& $0.63$	&		& \cite{Videler1993}\\
Bluefish				& $42$\q	& $5.3$		& $5.0\times 10^5$	& $0.028$	& $0.33$	& 			&		& \cite{Kayan1978}\\
Lake trout				& $21$\q	& $5.4$		& $2.0\times 10^5$	& $0.034$	& $0.33$	& 			&		& \cite{Kayan1978}\\
Largemouth bass			& $24.5$	& $6.4$		& $1.2\times 10^5$	& $0.050$	& $0.23$	& $0.68$	& $44$	& \cite{Jayne1995}\\
Pacific jack mackerel	& $27$\q	& $5.7$		& $5.0\times 10^5$	& $0.030$	& $0.31$	& 			&		& \cite{Hunter1971}\\
Rainbow trout\Juv\mi 		& $5.5$		& $8.8$		& $1.6\times 10^4$	& $0.184$	& $0.38$	& $0.57$	&		& \cite{Webb1984}\\
Rainbow trout\ma			& $56$\q	& $9.0$		& $2.5\times 10^5$	& $0.054$	& $0.25$	& $0.71$	&		& \cite{Webb1984}\\
Rainbow trout			& $20.1$	& $8.2$		& $1.1\times 10^5$	& $0.067$	& $0.26$	& $0.75$	& $47$	& \cite{Webb1988a}\\
Sockeye salmon			& $20.4$	& $10.4$	& $8.0\times 10^4$	& $0.097$	& $0.31$	& $0.60$	&		& \cite{Webb1973}\\
Yellowbelly rockcod	\Juv & $7.6$		& $22.2$	& $2.3\times 10^4$	& $0.385$	& $0.38$	& 			&		& \cite{Archer1989}\\
Yellowbelly rockcod		& $29$\q	& $10.5$	& $2.2\times 10^5$	& $0.065$	& $0.30$	& 			&		& \cite{Archer1989}\\[5pt]
							
\textbf{Other fishes}	&			&			&					&			&			&			&		& \\
Atlantic cod			& $25$\q	& $16.4$	& $1.2\times 10^5$	& $0.124$	& $0.30$	& $0.62$	&		& \cite{Webb2002}\\
Atlantic cod			& $63$\q	& $10.6$	& $3.1\times 10^5$	& $0.061$	& $0.28$	& 			&		& \cite{Webber2001}\\
Atlantic cod			& $49$\q	& $10.6$	& $3.7\times 10^5$	& $0.059$	& $0.25$	& $0.75$	&		& \cite{Videler1993, Videler1978}\\
Atlantic silverside		& $7.5$		& $10.9$	& $1.7\times 10^4$	& $0.224$	& $0.27$	& 			&		& \cite{Parrish1988}\\
Common bream			& $19$\q	& $3.4$		& $8.5\times 10^4$	& $0.031$	& $0.29$	& $0.76$	&		& \cite{Bainbridge1963}\\
Common dace				& $17.5$	& $4.9$		& $3.5\times 10^5$	& $0.027$	& $0.29$	& 			&		& \cite{Bainbridge1958}\\
Goldfish (Eggfish)		& $5.3$		& $10.6$	& $7.1\times 10^3$	& $0.538$	& $0.54$	& $0.41$	&		& \cite{Blake2009}\\
Goldfish (Fantail)		& $5.7$		& $10.4$	& $7.7\times 10^3$	& $0.512$	& $0.47$	& $0.53$	&		& \cite{Blake2009}\\
Goldfish (Common)		& $5.1$		& $4.7$		& $2.1\times 10^4$	& $0.093$	& $0.40$	& $0.80$	&		& \cite{Blake2009}\\
Goldfish (Comet)		& $5.7$		& $3.9$		& $2.3\times 10^4$	& $0.067$	& $0.44$	& $0.58$	&		& \cite{Blake2009}\\
Goldfish				& $18.8$	& $4.3$		& $1.5\times 10^5$	& $0.030$	& $0.30$	& 			&		& \cite{Bainbridge1958}\\
Lake sturgeon			& $15.7$	& $12.8$	& $4.1\times 10^4$	& $0.168$	& $0.48$	& $0.65$	& $46$	& \cite{Webb1986}\\
Mullet					& $27$\q	& $9.6$		& $3.0\times 10^5$	& $0.056$	& $0.33$	& 			&		& \cite{Kayan1978}\\
Saithe					& $36.4$	& $7.0$		& $3.9\times 10^5$	& $0.038$	& $0.23$	& $0.76$	&		& \cite{Hess1984}\\
Thinlip grey mullet		& $36$\q	& $7.7$		& $3.8\times 10^5$	& $0.042$	& $0.23$	& $0.76$	&		& \cite{Videler1993}\\
Thicklip grey mullet\Juv&$12.6$	& $6.4$	& $2.3\times 10^4$	& $0.113$	& $0.34$	& $0.70$	& $31$	& \cite{Muller2002}\\
Tiger musky				& $18.3$	& $9.4$		& $9.6\times 10^4$	& $0.081$	& $0.25$	& $0.60$	& $50$	& \cite{Webb1988a}\\
West African lungfish	& $55$\q	& $14.6$	& $7.0\times 10^3$	& $0.463$	& $1.02$	& 			&		& \cite{Horner2008}\\
West African lungfish	& $55$\q	& $14.6$	& $6.0\times 10^4$	& $0.158$	& $0.75$	& 			&		& \cite{Horner2008}\\[5pt]
\end{tabular}
\end{center}
\end{table}

\begin{table}
\label{tab:3}
\caption{see Table 1.}
\begin{center}
\begin{tabular}{@{}lrr cc cc cc@{}}
\hline\\[-5pt]
Species 				& $L$ (cm)	& $\SR$		& $\Reyn$			& $\Li$ 	& $\St$ 	& $U/V$		& $\A$ (deg)& Sources \\[5pt] \hline\\[-5pt]
														
\textbf{Elongated}		&			&			&					&			&			&			&		& \\										
Atlantic needlefish		& $23$\q	& $21.7$	& $1.2\times 10^5$	& $0.167$	& $0.34$	& $0.69$	&		& \cite{Liao2002}\\
American eel			& $21$\q	& $25.8$	& $6.0\times 10^4$	& $0.280$	& $0.31$	& $0.73$	&		& \cite{Tytell2004}\\
American eel			& $36$\q	& $25.8$	& $1.3\times 10^5$	& $0.192$	& $0.37$	& $0.79$	&		& \cite{Gillis1998}\\
European eel			& $22$\q	& $27.9$	& $4.0\times 10^4$	& $0.369$	& $0.48$	& $0.60$	&		& \cite{DAout1999}\\
European eel			& $73$\q	& $27.9$	& $2.5\times 10^5$	& $0.167$	& $0.52$	& 			&		& \cite{Ellerby2001}\\
Garfish					& $44$\q	& $18.0$	& $4.0\times 10^5$	& $0.098$	& $0.34$	& 			&		& \cite{Kayan1978}\\
Great sand-eel			& $30$\q	& $19.2$	& $1.2\times 10^5$	& $0.148$	& $0.31$	& $0.67$	&		& \cite{Videler1993}\\
Hagfish					& $31$\q	& $33.2$	& $6.4\times 10^4$	& $0.347$	& $0.56$	& $0.49$	&		& \cite{Long2002}\\
Lesser sand-eel			& $9.0$		& $16.4$	& $2.2\times 10^4$	& $0.296$	& $0.41$	& $0.64$	&		& \cite{Videler1993}\\
Longnose gar			& $57$\q	& $21.9$	& $3.3\times 10^5$	& $0.125$	& $0.59$	& $0.67$	&		& \cite{Long1996}\\[5pt]
							
\textbf{Others}			&			&			&					&			&			&			&		& \\
Axolotl					& $17.7$	& $19.2$	& $4.4\times 10^4$	& $0.242$	& $0.57$	& $0.59$	&		& \cite{DAout1997}\\
Bullfrog tadpole\Juv	& $4.7$		& $11.0$	& $2.3\times 10^4$	& $0.193$	& $0.79$	& $0.58$	&		& \cite{Wassersug1985}\\
Green frog tadpole\Juv	& $5.0$		& $10.6$	& $2.1\times 10^4$	& $0.195$	& $0.60$	& $0.60$	&		& \cite{Wassersug1985}\\
Lesser siren			& $34$\q	& $31.3$	& $1.7\times 10^5$	& $0.202$	& $0.54$	& $0.61$	&		& \cite{Gillis1997}\\
Medicinal leech			& $10.0$	& $19.5$	& $1.8\times 10^4$	& $0.384$	& $0.63$	& $0.70$	&		& \cite{Jordan1998}\\
Saltwater crocodile		& $93$\q	& $62.5$	& $4.2\times 10^5$	& $0.339$	& $0.78$	& 			&		& \cite{Seebacher2003}\\
\end{tabular}
\end{center}
\end{table}
\baselineskip12pt

\end{document}